\documentclass[12pt, reqno]{amsart}

\newtheorem{theorem}{Theorem}[section]
\newtheorem{lemma}[theorem]{Lemma}
\newtheorem{corollary}[theorem]{Corollary}
\theoremstyle{definition}
\newtheorem{definition}[theorem]{Definition}

\newtheorem{remark}[theorem]{Remark}

\numberwithin{equation}{section}

\usepackage{epsfig}
\usepackage{rotating}
\usepackage{url}
\usepackage{amsfonts}
\usepackage{mathrsfs}
\usepackage{textcomp}
\usepackage{bm}
\usepackage[round]{natbib} 
\usepackage[OT1]{fontenc}
\usepackage{graphicx,epsfig}
\usepackage{epstopdf}
\usepackage{amscd}
\usepackage{amsmath,latexsym,amssymb,amsthm}
\begin{document}

\title[Change in mean of functional data]{On existence of a change in mean of functional data}

%    Remove any unused author tags.

%    author one information
\maketitle
\begin{center}
\author{Buddhananda Banerjee$^{0a}$ and Satyaki Mazumder $^{1b}$}\\
\address{Department of Mathematics \& Statistics \\ Indian Institute of Science Education and Research, Kolkata$^{0,1}$ }\\

\email{buddha.banerjee@iiserkol.ac.in$^a$, satyaki@iiserkol.ac.in$^b$}
\end{center}

\footnotetext{
 corresponding author\/: buddha.banerjee@iiserkol.ac.in.}

\begin{abstract}
	Functional data often arise as  sequential temporal observations over a continuous state-space. A set of functional data with a possible change in its structure  may lead to a wrong conclusion  if it is not taken in to account. So, sometimes,  it is crucial to know  about the existence  of change point in a given sequence of functional data before doing any further statistical inference.  
	We develop a new methodology to provide a test for detecting  a change in the mean function of the corresponding data.    To obtain the test statistic we provide an alternative estimator of the covariance kernel. The proposed estimator is asymptotically unbiased under the null hypothesis and, at the same time,  has smaller amount of  bias than that of the existing estimator.  We show here that under the null hypothesis the proposed test statistic is pivotal asymptotically.  Moreover, it is shown that under alternative hypothesis the test is consistent  for  large enough sample size.  It is also found that the proposed test is more powerful than the available test procedure in the literature.  From the extensive simulation studies we observe that  the proposed test outperforms the existing one with a wide margin in power for moderate sample size. The developed methodology performs satisfactorily for the average daily temperature of central England and monthly global average anomaly of temperatures.
	%------------------------

% \vspace{.05cm}

\end{abstract}

\section{Introduction} \label{sec1}
Functional data analysis (FDA) is becoming increasingly popular because of its wide applicability in
 various fields of statistics. The natural proximity of functional data to feature some real life
  observations is more appealing over its finite dimensional  representation and at the same time it is
   often noticed that FDA leads to more accurate inference in this regard. 
  \cite{ramsey_2005book} has enriched  the literature with  a detailed discussions on
    several techniques and usefulness of  FDA. Some recent developments in many more aspects  of
    FDA can be found in \cite{Ferraty_2011book}.  However, the inference and especially the prediction may alter
    if there exists an inherent change in the stochastic structure of the functional data observed
     temporally.  The change may occur at a unknown point of time  with in the chronological
     sequence  of data  but it is always challenging to test, statistically, whether the change has
occurred or not. For the cases of scalar and vector data a considerable amount of contributions can
 be found  from the works by  \cite{Cobb_1978}, \cite{Inclan_1994}, \cite{Davis_1995}, \cite{Antoch_1997}, \cite{Hovarth_1999}, \cite{Kokoszka_2000}, \cite{Kirch_2014} and references therein, among many others.
    In the context of functional data a change may occur in the mean function or in the covariance
     kernel of the  data or both. This paper shades light  on the discussion about the change in the
      mean function in particular. Recently, \cite{Berkes_2009} and \cite{Aue_2009} have proposed
       a method for detecting changes in the mean functions of an observed set of  functional data.  
   \cite{Berkes_2009}, in their pioneering work in this context,  have provided an  elegant test
    procedure to decide the existence of a significant amount of change in   the mean function,
    whereas \cite{Aue_2009} following  the method of \cite{Berkes_2009} have dealt with the
    detection of the position of the change in the mean function. In practice, both are  equally
     important to judge  whether there is a change in the mean function of the data and if there  is a
      significant  change at all then   detecting the location of it. For example, while analyzing the
      temperature of a certain  region over a long period of time, it is very important to 
       environmentalist to identify the time  point after which a significant change in the mean
        temperature is observed as a possible   effect of global warming. 
        %Now a days, leading scientists
        % of different realms are concerned about  levels of ozone layers, which is getting thinner above 
%          different locations around the globe. It is highly beneficial for them to know whether there is a
%    significant change in the  atmospheric ozone layers  at  a  particular location and if there is at all a
%      change, what has been  the possible time is  of  that change. 
     %Berkes et.\ al (2009) has shown that FDA has certain advantages over multivariate data analysis %while dealing with this kind of environment data. 
     In this paper we come up with a different methodology to analyze the functional  data subject to a possible change point  and  propose a new statistical test, which is more powerful than the existing one(s), for detecting the presence of a change in the mean function of the data. 
     Here we show  that under the null hypothesis, i.\ e.\ with no change in the  data, the proposed test
      statistic converges in distribution to a functional of  the Brownian bridges, as shown in \cite{Berkes_2009}. Moreover, we prove here that the test  is consistent under alternative hypothesis when
       the number of the observations becomes large enough. We provide an estimator of the covariance kernel which not only enjoys its property of consistency under the null hypothesis but also has less asymptotic bias compared to that of the estimator  provided by \cite{Berkes_2009} or \cite{Aue_2009} under the alternative hypothesis. Because of the reduction in the asymptotic bias while estimating the  covariance kernel, we successfully  obtain  that the proposed test has better power than the existing method by  \cite{Berkes_2009}.  The outcomes of an extensive simulation study reflects the same. It is also noted that our method outperforms the existing method in a wide margin   for  small samples. Therefore, it is more advantageous to use the proposed  method in practice for deciding with the presence of significant change in the mean of the functional data, specially when the data size is not big enough. 
\par 
The organization of the paper is as follows. In Section \ref{sec2}, we introduce the required notation and definitions for introducing the subject. The details of the model, discussed in the paper, are described in this section. Section \ref{sec3} deals with the testing methodology and main results of the paper. In this section  we provide the theorems about the consistency of the proposed estimator of covariance kernel, asymptotic null distribution and asymptotic consistency of the test procedure. In Section \ref{sec4} simulations results are provided in great detail where we show that our method substantially improves over the  existing method in terms of power of the test. In Section \ref{sec5} we show the performance of our test in real data. Remarks and conclusion of the  work are given in the Section \ref{sec6}. Finally we provide  the required proofs of the  results of section  \ref{sec3} in the Appendix (Section \ref{sec7}).   
\section{Preliminaries  and assumptions} \label{sec2}
Let , $X_{i}(t)$ for $i=1,\ldots,N$, be Hilbert-valued random functions defined over a compact set $\tau=[0,1]$. We assume that $X_{i}s$ are independent. We are interested to  check the  equality of the mean functions of  $X_i$ for all $i=1,2,\cdots,N$. More precisely,  the null hypothesis  to  test will be 
\[
H_0: E(X_{1}(t)) = E(X_{2}(t)) = \cdots = E(X_{N}(t)).
\] 
It is important to note  that nothing is presumed  about any property of the  common mean   under the null hypothesis.  
\par
Under alternative hypothesis we assume that the null hypothesis $H_0$ does not hold.
We deal with the  situation when the data contains  at most  one change point, however, in case of applications we elaborate  how to implement this method with multiple  change points case. In particular, in Section \ref{sec5}, we specifically deal with the situation with more than one  change points. 
 There  the data can be subdivided into several consecutive parts and within each part the mean function remains constant but it deviates between different contiguous parts. 
The details of the model with single change point is discussed in the sub-Section \ref{subsec1}. 

\par 
Under the null hypothesis we express $X_{i}$, $i=1,\ldots, N$, in the following manner.
\begin{align}
\label{eq1:H0 X to Y}
& X_{i}(t)  = \mu(t) + Y_{i}(t) \notag
\\
& E(Y_{i}(t)) = 0.
\end{align}
Now we specify the assumptions about mean function $\mu$ and random element $Y_{i}$, based on  which the  asymptotic behaviour of the test statistic can be determined. From here on words all integrations are computed over the compact set $\tau$, unless otherwise mentioned.  
\subsection{Assumptions}\label{subsec1}
\begin{itemize}
\item[A1.] The mean function is square integrable that is, $\mu \in L^2(\tau)$, and the  unobservable random component $Y_{i}s$, are independent and identically distributed random elements in $L^2(\tau)$ with 
\[
E(Y_{i} (t)) = 0  ~~~~~~~~~~~~~~~~~~~~~~~\forall t\in \tau,
\]
for $i=1,\ldots,N$ and 
\begin{align}
\label{eq2:square integrability of Y}
E||Y_i||^2 = \int E(Y_{i}^2(t))\, dt < \infty.
\end{align}
The covariance kernel is defined  as 
\begin{align}
\label{eq3: c_t,s}
c(t,s) = E(Y_{i}(t)Y_{i}(s)) \,\,\,\,\,\,\,\,\, t,s \in \tau
\end{align}
with the  assumption  that $c(t,s)$ $\in$ $L^2(\tau\times \tau)$. 
Assumption 1 implies that the covariance operator of $Y$, which is  a positive definite symmetric Hilbert-Schmidt (H-S) operator mapping from $L^2(\tau)$ to itself,  will be of the form
\begin{align}
\label{eq4: cov op of Y}
C(x) = E[\langle Y,x \rangle Y].
\end{align}
%%(Here we use the assumption that $Y_i$, $i=1,\ldots, N$ are identically distributed random
% elements as $Y$ in $L^2(\tau)$). 
The evaluation of $C(x)$ at $t$, i.e., $C(x)(t)$, is given by
\[
C(x)(t) = \int c(t,s)x(s)\, ds \,\,\,\,\,\,\,\,\, \forall t \in \tau.
\]
Moreover, Mercer's theorem in \cite[][Chapter 4]{Indritz_1963} implies that $c(t,s)$ has the following spectral decomposition\/:
\begin{align}
\label{eq5:spec decom of c_t,s}
c(t,s) = \sum_{l=1}^{\infty} \lambda^{l} \upsilon^{l}(t)\upsilon^{l}(s) \,\,\,\,\,\,\,\, t,s \in \tau,
\end{align}
where each real scalar $\lambda^{l}$ and function $\upsilon^{l}$ (in $L^2(\tau)$) are defined, for $t\in \tau$, as
\begin{align}
\label{eq6:eigvalue and eigfunction of c_t,s}
C(\upsilon^{l})(t) &= \lambda^{l} \upsilon^{l}(t), ~ l = 1,2\ldots, \notag
\\
\mbox{i.e.}\ , \int c(t,s)\upsilon^{l}(s) &= \lambda^{l} \upsilon^{l}(t), ~ l = 1,2,\ldots.
\end{align}
In other words, $\lambda^{l}s$ and $\upsilon^{l}s$ are the eigenvalues and the corresponding eigenfunctions respectively, of the operator $C( . )$. Since the eigenfunctions of the positive definite symmetric operator, $C( . )$, form a complete orthonormal basis of $L^2(\tau)$ and eigenvalues are positive, Karhunen-Lo$\acute{\mbox{e}}$ve representation of $Y_{i}$ holds good in $L^2(\tau)$ and is given by
\begin{equation}
\label{eq7: represenation of Y}
Y_{i}(t) = \sum_{l=1}^{\infty} \sqrt{\lambda^{l}} \delta_{i}^{l} \upsilon^{l}(t),
\end{equation}
where $\sqrt{\lambda^{l}} \delta_{i}^{l}$ = $\langle Y_{i}, \upsilon^{l}\rangle$ = $\int Y_{i}(s)\upsilon^{l}(s)$ is known as $l$th functional principal component score. By construction, the elements of the sequence $\{\delta_{i}^{l}\}^{l}$ are uncorrelated random variables with zero mean and unit variance and $\{\delta_{i}^{l}\}^{l}$ and $\{\delta_{j}^{l}\}^{l}$ are independent  for $i\neq j$. 
\item[A2.] There exists some positive integer $d$, such that the eigenvalues $\lambda^{l}$ satisfy
$$
\lambda^{1}>\lambda^{2}>\ldots>\lambda^{d}>\lambda^{d+1}.
$$
\item[A3.] $Y_{i}$, $i=1,\ldots,N$, satisfy
\[
E(||Y_{i}||^4) = \int E(Y_{i}(t))^4 \, dt <\infty.
\] 
\par 
%We provide the proof of the consistency of the testing procedure under the alternative of single %change point present in the data. Consistency of testing procedure under multiple change point %can be shown with much more technical difficulties following the same line as done in section %(\ref{sec3}). However, those technical difficulties will break the flow of the paper. We empirically %study the performance of the testing procedure under multiple change points in section $(\ref{sec6}). The model under the alternative of single change point is formulated in next %assumption.
\item[A4.] Under the alternative, with an existence of   single change point the observations, $X_i$, $i=1,\ldots, N$ can be represented as follows
\begin{align}
\label{eq8: HA X to Y}
X_{i}(t) = \begin{cases}
         \mu_1(t) + Y_{i}(t), & 1\leq i\leq k^* \\ 
         \mu_2(t) + Y_{i}(t), & k^*<i\leq N
           \end{cases}
\end{align}
where $Y_{i}$, $i=1,\ldots, N$  satisfy the assumption A1, $\mu_{j}(t)$, $j=1,2$ are in $L^2(\tau)$ and $k^*$ = $[N\theta]$, with $\theta\in (0,1)$. Therefore, we assume that under the alternative hypothesis of single change point a  change  may occur in the mean function but the covariance kernel remains the same before and after the change in the  data. Keeping this in consideration we estimate the covariance kernel in the following section and develop a new methodology to test $H_0$.  
\end{itemize}
 
%-------------------------------------------------- 
\section{Methodology and Main results} \label{sec3}
%--------------------------------------------------
To estimate the covariance kernel let us define  the piecewise sample means  for two segments 
\begin{align}
\label{eq9: mu_1_hat}
\widehat{\mu}_k(t)& =\displaystyle\frac{1}{k}\sum_{i=1}^{k}X_i(t), \\
\label{eq10: mu_2 hat}
\widetilde{\mu}_k(t)& =\displaystyle\frac{1}{N-k}\sum_{i=k+1}^{N}X_i(t),
\end{align}
where $k$ = $[Nu]$ with $u$ $\in$ $(0,1)$, implying $1\leq k< N$. For $u=1$ we define $\widehat{\mu}_N(t) = \displaystyle \frac{1}{N} \sum_{i=1}^{N}X_{i}(t)$. With the help of equations (\ref{eq9: mu_1_hat}) and (\ref{eq10: mu_2 hat}), the newly proposed estimator  of covariance kernel is
\begin{small}

\begin{align}
\label{eq11: c_u(t,s)}
\widehat c_u(t,s) = \frac{1}{N}\left[\sum_{i=1}^{k}\left(X_i(t)-\widehat{\mu}_k(t)\right)\left(X_i(s)-\widehat{\mu}_k(s)\right)
+\sum_{i=k+1}^{N}\left(X_i(t)-\widetilde{\mu}_k(t)\right)\left(X_i(s)-\widetilde{\mu}_k(s)\right)
\right].
\end{align}
\end{small}
For $u=1$, we define 
$\widehat{c_{1}}(t,s) = \frac{1}{N}\left[\sum_{i=1}^N \left(X_i(t)-\widehat{\mu}_N(t)\right)\left(X_i(s)-\widehat{\mu}_N(s)\right)\right]$,
which is commonly used as estimator of covariance kernel, see for example, \cite{Berkes_2009} and \cite{Aue_2009}. With the newly proposed estimator  of the covariance kernel we obtain  the most important finding of this paper which is narrated in the following theorem.
\begin{theorem}
\label{thm1}
Defining $c_u(t,s):=c(t,s)+\theta(1-\theta)\Delta(t)\Delta(s)~f_{\theta}(u),$ under assumption $\mbox{\textit{A}}4$, 
\[
\int \int [\widehat c_u(t,s)- c_u(t,s)]^2\, dt ds \xrightarrow{P} 0,  \mbox{as}~ N\uparrow \infty,
\]
%\\ $$\widehat c_u(t,s)\xrightarrow{P} c_u(t,s)$$   in $L^2(\tau\times \tau)$ as $N\uparrow \infty$, 
%in the sense that 
where, $$f_{\theta}(u)=\frac{\max\{u,\theta\}-\min\{u,\theta\}}{\max\{u,\theta\}(1-\min\{u,\theta\})}\in [0,1]$$
with $\theta \in (0,1), u\in(0,1]$ and $\Delta(t)=\mu_1(t)-\mu_2(t)$.
\end{theorem} 
\noindent
Proof\/: The proof of the theorem is provided in the  Appendix,  \ref{sec7}. \hfill $\square$
\par 
\begin{corollary}
 If null hypothesis is true then $\widehat c_u(t,s)\xrightarrow{P} c(t,s)$ for all $u\in (0,1]$.
\end{corollary}
Some more interesting observations, which show the greater  applicability of  Theorem \ref{thm1}, are immediate  from it.  
\begin{remark}
It can be easily checked that $c_{u}(t,s)$ is a positive definite, symmetric satisfying
\[
\int \int c_{u}^2(t,s)\,dt ds < \infty,
\]
and hence is a covariance kernel.
\end{remark}
\begin{remark}
If $u=1$, that is, if commonly used estimator of $c(t,s)$ is used then it is readily observable  that, under alternative, $\widehat c_1(t,s)\xrightarrow{P} c(t,s)+\theta(1-\theta)\Delta(t)\Delta(s)$ = $\widetilde{c}(t,s)$, say, which is also proved by \cite{Berkes_2009}. We note here that whenever $H_0$ is false, $\widehat c_1(t,s)$ has a constant bias $\theta(1-\theta)\Delta(t)\Delta(s)$. Therefore, for any $u\in (0,1)$, the asymptotic bias of the estimator $\widehat c_u(t,s)$ is less than that of $\widehat c_1(t,s)$ under alternative hypothesis. 
\end{remark}
\begin{remark}
If $u=\theta$, that is, when the data is partitioned in true position, then $\widehat c_\theta(t,s)\xrightarrow{P} c(t,s)$ and in that case asymptotic bias of $\widehat c_\theta(t,s)$ is zero whereas asymptotic bias of $\widehat c_1(t,s)$ remains $\theta(1-\theta)\Delta(t)\Delta(s)$. 
\end{remark}

A few more notations and definitions are needed to be introduced here to state the further results
\begin{definition}
\label{def1}
The orthonormal functions $\omega_{u}^{l}(t)$ in $L^2(\tau)$ corresponding to real scalars $\gamma_{u}^{l}$ are defined as orthonormal eigenfunctions corresponding to eigenvalues $\gamma_{u}^{l}$ of the covariance operator $C_{u}( . )$ from $L^2(\tau)$ to $L^2(\tau)$, defined as $C_u(x)(t)$ = $\int c_u(t,s)x(s)\,ds$, satisfying the relation
\begin{equation}
\label{eq12: eigenvalue and eigenfunctions of c_u}
\int c_u(t,s) \omega_{u}^{l} (s)\, ds = \gamma_{u}^l \omega_{u}^{l}(t) 
\end{equation}
\end{definition}
\begin{definition}
\label{def2}
The estimates of the eigenvalues $\gamma_{u}^{l}$ and $\omega_{u}^{l}$ are denoted  as $\widehat{\lambda}_{u}^{l}$ and $\widehat{\upsilon}_{u}^{l}$, satisfying the relation
\begin{equation}
\label{eq13: eigenvalue and functions of hat_c_u}
\int \widehat{c}_u(t,s) \widehat{\upsilon}_{u}^{l} (s)\, ds = \widehat{\lambda}_{u}^l \widehat{\upsilon}_{u}^{l}(t). 
\end{equation}
\end{definition}
With the above two definitions we have the following important observations can be noted
\begin{corollary}
\label{cor1}
Under the assumption A4, for every $1\leq l \leq d$ and $u\in (0,1]$, we have 
\begin{eqnarray}
&\widehat{\lambda}_{u}^{l}  \xrightarrow{P} \gamma_{u}^l&\\
&\displaystyle\int [\widehat{\upsilon}_{u}^{l}(t)-\widehat{c}_u^{l}\,\omega^{l}_{u}(t)]^2\, dt  \xrightarrow{P} 0,&
\end{eqnarray}
where $\widehat{c}_u^{l}$ = sgn$\langle\omega^{l}_{u}, \widehat{\upsilon}_{u}^{l}\rangle$.
\end{corollary}
\noindent
Proof\/: The proof follows from the Theorem \ref{thm1} and  lemmas 4.2 and 4.3 of  \cite{Bosq_2000}. \hfill $\square$
\begin{remark}
Under $H_0$, for all $1\leq l\leq d$ and $u\in (0,1]$, $\widehat{\lambda}_{u}^{l}\xrightarrow{P} \lambda^{l}$ and $\widehat{\upsilon}_{u}^{l}$ converges to $\upsilon^{l}$ in probability, in $L^2(\tau)$. Moreover, under alternative hypothesis, if $u=\theta$ then for all $1\leq l\leq d$, $\widehat{\lambda}_{\theta}^{l}\xrightarrow{P} \lambda^{l}$ and $\widehat{\upsilon}_{\theta}^{l}$ converges to $\upsilon^{l}$ in probability, in $L^2(\tau)$. It, in fact,  can be easily seen that 
$$\displaystyle\sup_{0<u
	\leq 1}\int [\widehat{\upsilon}_{u}^{l}(t)-\widehat{c}_u^{l}\,\upsilon^{l}_{u}(t)]^2\, dt  \xrightarrow{P} 0,$$
\end{remark}

\par 
In the direction of the eigenfunctions $\widehat{\upsilon}^{l}_{u}$ corresponding to the largest $d$ eigenvalues $\widehat{\lambda}^{l}_{u}$  the noncentral scores can be obtained  as 
\begin{align}
\label{eq14: scores}
\widehat{\eta}_{i,l}(u) = \int X_{i}(t) \widehat{\upsilon}^{l}_{u}(t) \, dt, ~ i=1,\ldots,N, ~ l=1,\ldots d.
\end{align}
Utilizing the score functions, as defined  above,  we provide a statistic and its distributional convergence in the following theorem which will important to know to construct  the test statistic and perform the asymptotic test.  First we define  the statistic based on the self normalized partial sums in $d$ dimensions 
\begin{align}
\label{eq15: statistic T}
R_{N}(u) = \frac{1}{N} \sum_{l=1}^{d} \frac{1}{\widehat{\lambda}^{l}_{u}} \left(\sum_{i=1}^{[Nu]}\hat\eta_{i,l}(u)-u\sum_{i=1}^{N}\hat\eta_{i,l}(u)\right)^2
\end{align}
Further denoting  $B_1(\cdot),\ldots,B_{d}(\cdot)$ be the  standard independent Brownian bridges, the theorem is provided
\begin{theorem}
\label{thm2}
Let the assumptions A1 to A3 hold. Then with the proper embedding of  Skorohod topology in $D[0, 1]$, under $H_0$ 
\begin{align}
\label{eq16: convergence under H0}
R_N(u)\xrightarrow{d}\displaystyle \sum_{l=1}^{d}B_l^2(u), \qquad 0\leq u\leq 1 . 
\end{align} 
\end{theorem}
\noindent
Proof\/: Proof of the theorem is given in Appendix,  \ref{sec7}. \hfill $\square$
\par 
Finally we define the test statistic as follows\/:
\begin{align}
\label{eq17: Test Stat}
H_{N,d} := \frac{1}{N^2}\sum_{l=1}^d\sum_{[Nu]=1}^N\frac{1}{\hat{\lambda}^{l}_u}\left(\sum_{i=1}^{[Nu]}\hat\eta_{i,l}(u)-u\sum_{i=1}^{N}\hat\eta_{i,l}(u)\right)^2.
\end{align}
Using the   Theorem \ref{thm2} it is immediate to see that $H_{N,d}\xrightarrow{d} \int_{\tau}\sum_{l=1}^{d}B_l^2(u) du$  under $H_0$,
because integral is a continuous functional and 
$ U(R_{N}(\cdot))\xrightarrow{d} \displaystyle U\left(\sum_{l=1}^{d}B_l^2(\cdot)\right)$ for any continuous functional $U\/: D[0,1]\rightarrow \Re $ (see \cite{Berkes_2009} for further details). The distribution of the limiting random variable can be found in \cite{kiefer_1959} and its $(1-\alpha)$th quantile are given in Table 1 of \cite{Berkes_2009}. We use this asymptotic critical values for performing the tests and  $H_0$ is rejected at 100$(1-\alpha)\%$ confidence level if the observed value of $H_{N,d}$ is bigger than the tabulated $(1-\alpha)$th quantile $K_d(\alpha)$ in \cite{Berkes_2009} . 
\par 
Now we  show that the proposed  test is consistent under the alternative hypothesis. Basically we show here that $H_{N,d}\xrightarrow{P} \infty$ under the hypothesis of single change point. The following theorem assures the claim.
\begin{theorem}
\label{thm3}
\textnormal{Under the assumption $\mbox{A}4$, 
\[
\frac{1}{N}H_{N,d}\xrightarrow{P}\displaystyle\sum_{l=1}^{d} \int_{0}^{1}\frac{g_l^2(u)}{\gamma_u^l} du,
\]
where $g_l(u)=\min\{\theta,u\}\left(1-\max\{\theta,u\}\right)\int_{\tau}\Delta (t) \omega_{u}^{l}(t)dt$}. 
\end{theorem}
\noindent
Proof\/: The proof follows from  Theorem \ref{thm1}  and the following lemma.
\hfill $\square$
\begin{lemma}
\label{lemm1}
\textnormal{Under the assumption A4,
$\displaystyle \sup_{0 \leq u \leq1} \left|N^{-1}R_N(u)-\displaystyle\sum_{l=1}^{d} \frac{g_l^2(u)}{\gamma_u^l} \right|=o_P(1)$}.
\end{lemma}
	%where 	$g_l(u)=\min\{\theta,u\}(1-\max\{\theta,u\})\int_{\tau}\Delta (t) \omega_l(t)dt$}
\noindent
Proof\/: Proof of the lemma follows from the proof of the Theorem 2 of \cite{Berkes_2009}. \hfill $\square$

Clearly from Theorem \ref{thm3} if $\int_{0}^{1}\frac{g_l^2(u)}{\gamma_u^l} du > 0$ for some $1\leq l\leq d$, then $H_{N,d}\xrightarrow{P} \infty$. 

Similar to \cite{Berkes_2009}, the change point $\theta$ is estimated by finding the value of $u$ which maximizes the function $R_{N}(u)$. For uniqueness we define the estimator formally as
\begin{align}
\label{eq18:estimate of theta}
\widehat{\theta}_{N} = \mbox{inf} \{u': R_{N}(u')=\sup_{0\leq u\leq 1}R_{N}(u)\}. 
\end{align}
It can be easily shown that (using lemma \ref{lemm1}), under the assumption A4, $\widehat{\theta}_{N}\xrightarrow{P} \theta$ provided $<\Delta,\omega^{l}_{u}>\neq 0$ for all $u\in (0,1]$ (see for example the proposition 1 and its proof of \cite{Berkes_2009}).

\section{Simulation studies} \label{sec4}
\begin{figure}[t]
\centering
\includegraphics[width=10cm, height=14cm,angle=-90]{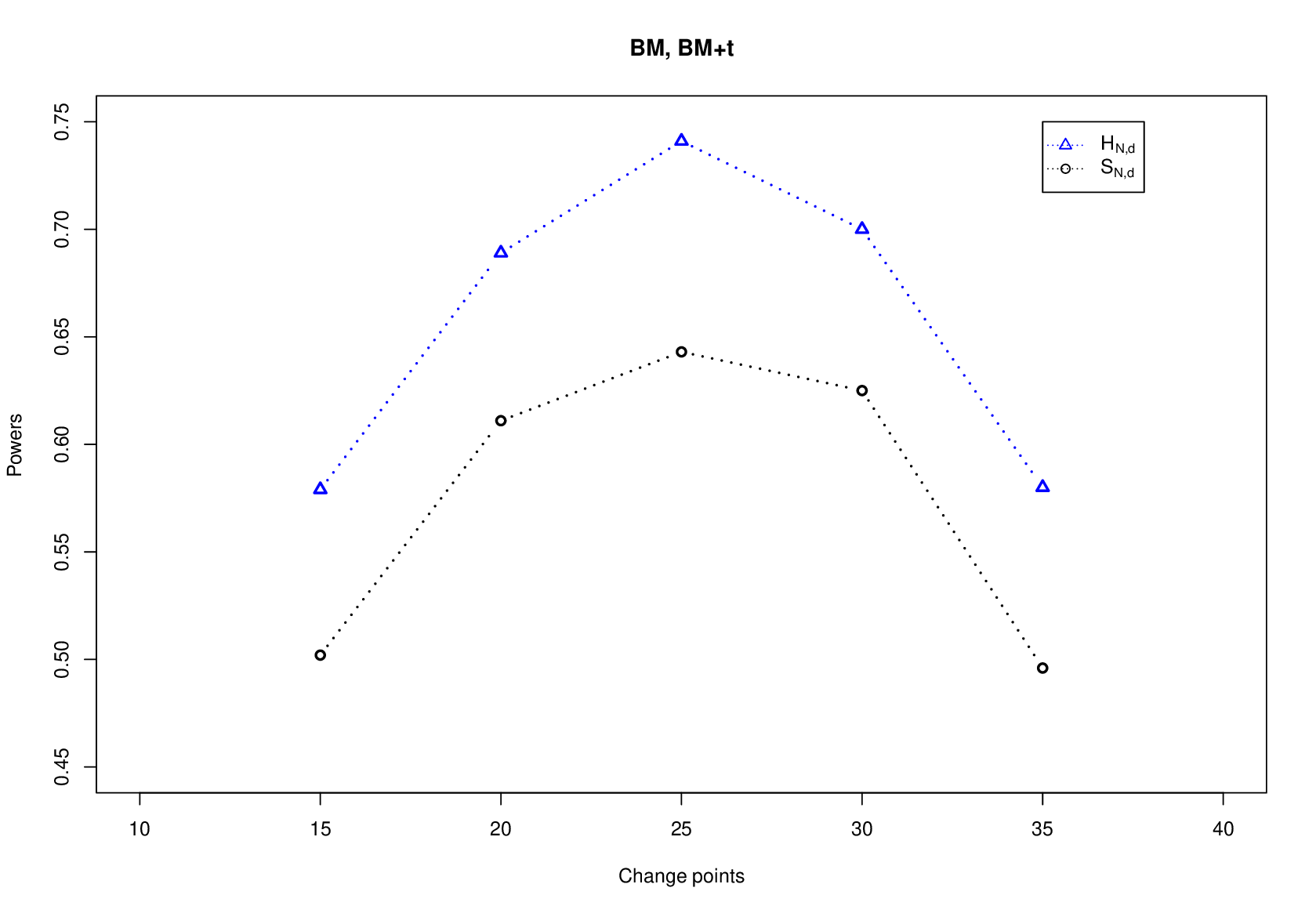}

\caption{{\small Power comparison of $H_{n,d}$ and $S_{n,d}$ for $N=50$ and $d=3$ with $\Delta(t)=t.$}}

\label{fig:50t}
\end{figure}

\begin{figure}[b]
	\centering
	\includegraphics[width=9cm, height=14cm,angle=-90]{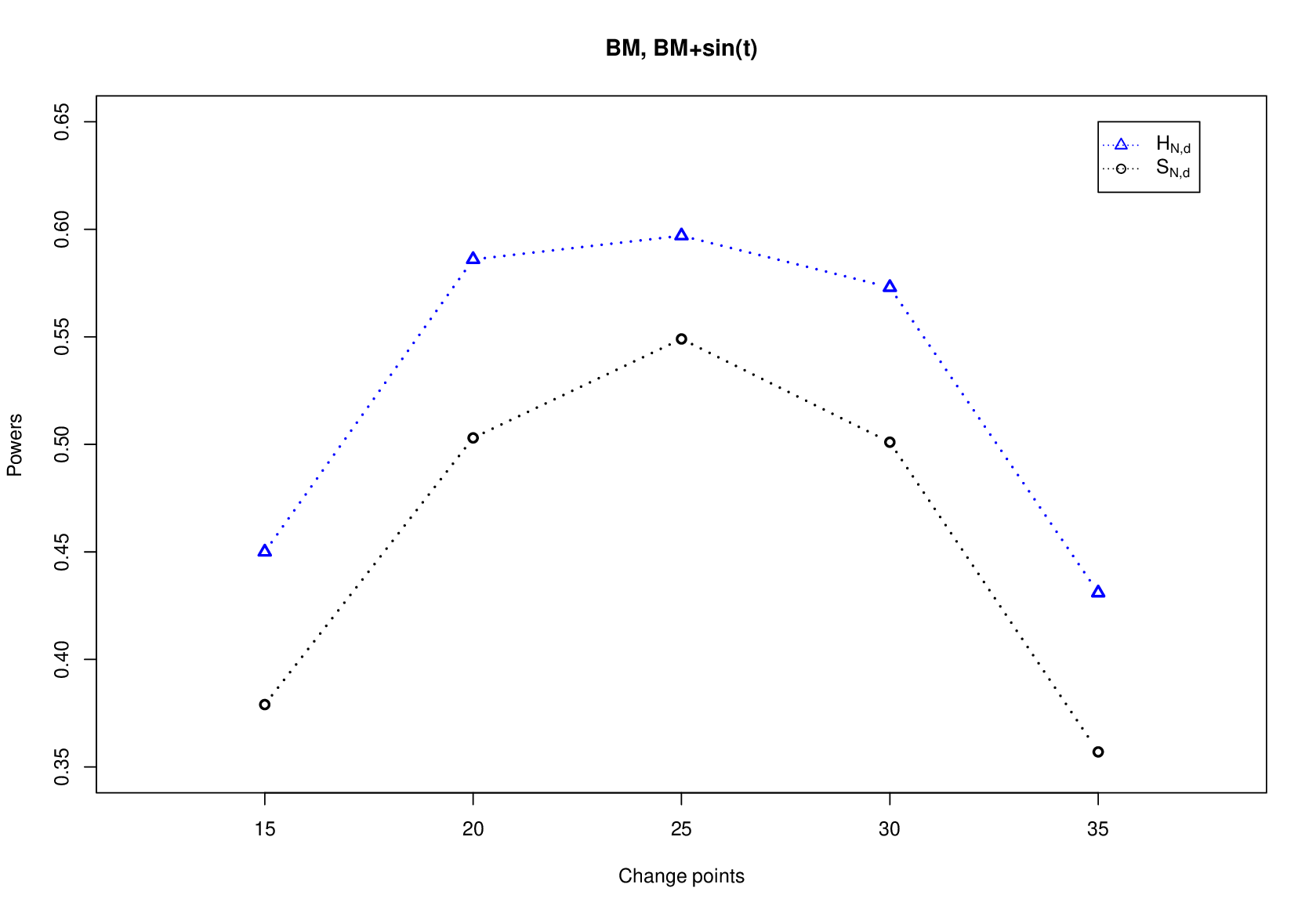}
	\caption{{\small power comparison of $H_{n,d}$ and $S_{n,d}$ for $N=50$ and $d=3$ with $\Delta(t)=\sin (t).$}}
	
	\label{fig:50sint}
\end{figure}

In this section we report a summery of the extensive simulation studies that we have conducted for moderate  and large sample sizes.  As proposed in  Section 3, we reject the null hypothesis  when the observed value of $H_{N,d}$ exceeds the corresponding critical  value $K_d(\alpha)$.  The critical values that are available in  \cite[][Table 1]{Berkes_2009}.  Without loss of generality initial mean function is considered to be zero.  For the first set of simulation studies the samples are generated from the standard Brownian motion (BM) over the interval $[0,1]$  and a drift of amount 
$t$ and $\sin(t)$ are considered after the presumed  locations of change point. 
The same is done for the standard Brownian bridge over $[0,1]$ and the mean shift after the change point is considered to be a  quadratic function $0.8t(1-t).$ To generate a sample from each of such Gaussian processes 1000 equidistant grid points are used. 750 Bspline basis functions are used to convert the grid data to functional data and first $3(=d)$ eigenfunctions are used to execute the testing procedures. For a pre-decided sample size and a specific change point the entire process is replicated 10000 times to assess the power of the test.  The considered  sample sizes $(N)$  are $50, 100, 150, 200, 300, 500.$ For any  particular sample size different possible locations of change  points ($k^*$) are chosen, to cover a wide range,  which are summarized  in   Table \ref{Table1} and  Figure \ref{fig:50t},  Figure \ref{fig:50sint}. 
For all practical purposes, we use the complete data together  for computing the estimated covariance kernel when $[Nu]$ = 1 or $[Nu]$ = $N-1$, otherwise as proposed in equation (\ref{eq11: c_u(t,s)})
\subsection{Small sample bias correction:} For small sample size (less than or equals to 100, say) we observe some fluctuations in the empirical size of the proposed test based on $H_{N,d}$.  To overcome this instability we propose a bias correction which helps us to get  empirical size reasonably close to $0.05$. Under the null, it is easy to observe that    
\begin{equation}
	E[\widehat c_u(t,s)]= \left(1-\frac{2}{N}\right)c(t,s). 
\end{equation}
So we suggest to multiply the correction factor with $\left(1-{2}/{N}\right)^{-1}$ with $\widehat c_u(t,s)$ to obtain the satisfactory results. Indeed  for the large sample the effect of the correction factor vanishes automatically and it hardy matters whether we use it or not. 
\subsection{Simulation findings\/:}
In all of the cases we find that the power curves for the proposed test based on $H_{N,d}$ strictly dominates that of the  $S_{N,d}$  proposed by  \cite{Berkes_2009}. For large sample (200 and above, say) the two power curves get very close to each other. But for small sample  we observe a remarkable gap between these two.  In particular, we provide the details of power for $N=100$ and $d=3$ at different point of change points starting from $ 15 $ to $85$ for Brownian motion and Brownian bridge in Table \ref{Table1}. We add two different functions, namely $t$, $\sin t$ with the mean of Brownian motion and add $0.8t(1-t)$ with the mean of standard Brownian bridge. In all of the above  cases it is found that the  proposed  method  has more power than that of the method by  \cite{Berkes_2009}  for all different locations of change points.  The Figure \ref{fig:50t} and the Figure \ref{fig:50sint} show the powers of two methods for sample size $50 (=N)$  at different point of changes, where the data have been simulated from standard Brownian motion and two different functions, $t$ and $\sin t$  are added separately with its mean at different locations of change for illustration purposes. It can be clearly observed that if sample size is small then our method is outperforming the method of \cite{Berkes_2009} with much larger difference. We also have done simulations  with different sample sizes and varieties of functions, e.\ g. $t^2$, $\sqrt{t}$, $\exp(t)$, $\cos(t)$ etc, being added to the mean of Brownian motion and Brownian bridge, and in all cases we have found that our method has a better power than that of existing method. 
This finding is quite intuitive because both test are asymptotic tests (both converging to the same asymptotic distribution) and the proposed one always has higher power than that of \cite{Berkes_2009}, mainly because the bias in the newly proposed estimate of covariance kernel under alternative is smaller than that in the usual estimate of covariance kernel used elsewhere. This satisfies the desirable quality of a better asymptotic test.  We also observe quite good performance of the  test statistic when the location of change point is $\leq N/4$ and  $\geq 3N/4$. 

%\textbf{we have to decide diagram and tables }

\section{Real Data analysis} \label{sec5}
 The findings of real data analysis to show the performance of proposed  test is demonstrated  in this section. Two temperature data have been analyzed. One data consists of average daily temperatures of central England for 228 years, from 1780 to 2007. The data has been taken from the website of British Atmospheric Data Centre. The second data, taken from Carbon Dioxide Information Analysis Center, consists of monthly global average anomaly of the temperatures from 1850 to 2012. Thus, these two data sets can be viewed as 228 curves with 365 measurements on each curve and 163 curves with 12 measurements on each curve, respectively. These two data sets are converted to functional data using 12 B-spline basis functions and 8 B-spline basis functions, respectively. Now  we discuss the performance of the  test statistics on these two  temperature data sets individually.
\par
\begin{figure}[t]
	\centering
	\includegraphics[width=9cm, height=14cm,angle=-90]{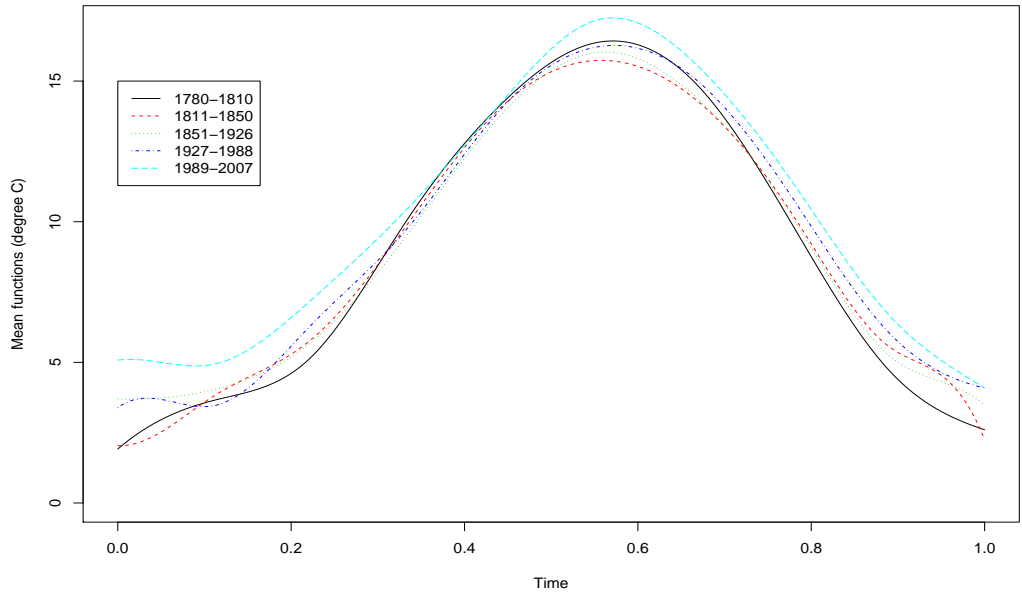}
	\caption{Segment wise mean functions of central England temperature  data }
	\label{fig:uk}
\end{figure}

 
To use the proposed test statistic for temperature data of  the central England we use first $8\, (=d)$ eigenfunctions explaining about $85\%$ of the total variability. Given the test indicates a change, the change point is estimated by calculating $\widehat{\theta}_{N}$ as described in the Lemma (\ref{eq18:estimate of theta}). Thereafter  dividing the data set into two parts the procedure is repeated  for the  each part until the test fails to reject the null hypothesis. The outcome of our method on this data has been provided in Table \ref{Table2}. It can be seen that the change points detected by our method and by the method of \cite{Berkes_2009} are very much adjacent. Both of the methods have detected 1850 and 1926 as possible change points. In case of other years of change point it is observed that the timings are very close, for example our method has detected a change in 1810 whereas \cite{Berkes_2009} has detected a change in 1808 and in the recent years our method has detected a change in 1989 and \cite{Berkes_2009} has detected 1993 as possible change point. Overall, it is important to note that both these methods have detected four change points in the given data. Table \ref{Table2} also shows the p-values corresponding to the observed value of the statistic for both of the methods. From the p-values it is noted that the  p-values of proposed test are much more smaller than the p-values of existing method showing the greater power of our test. The mean functions for each partitioned data sets are provided in the Figure \ref{fig:uk}. The picture clearly shows that there is a upward trained in the structure of the mean function from one period to other. 

%\textbf{Ekhane ebar oder ta te kon year detect korechilo r amader ta kon year korlo segulo bolte hobe}

\par
\begin{figure}[b]
	\centering
	\includegraphics[width=10cm, height=14cm,angle=-90]{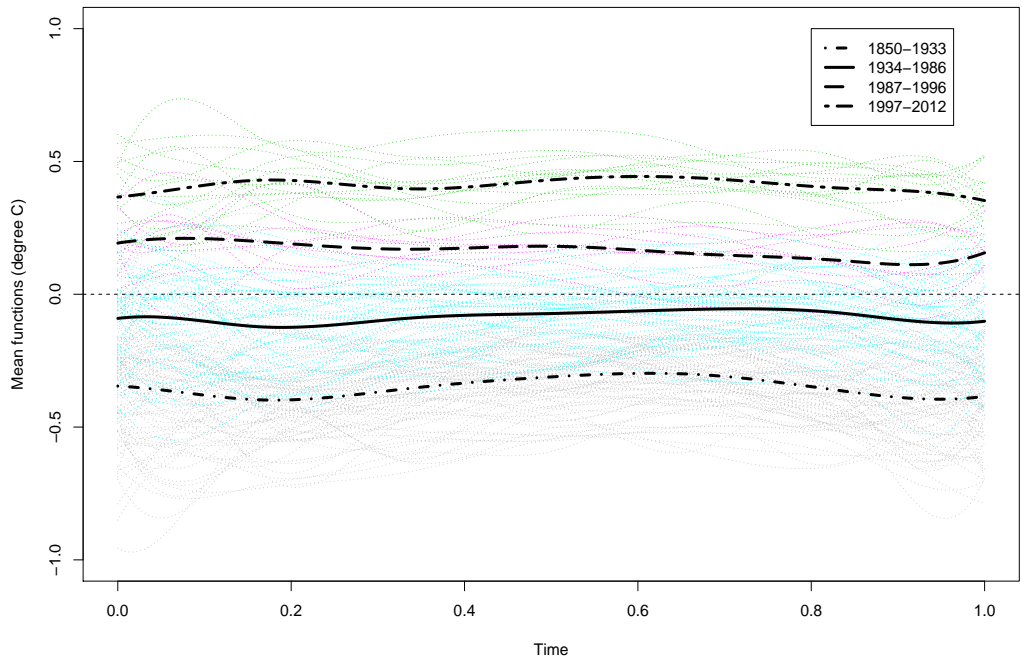}
	\caption{\small Segment wise mean functions of average  anomaly of global temperature  data }
	\label{fig:global}
\end{figure}

 
For the monthly average anomaly of the global  temperature data of 163 years, first  $3\, (=d)$ eigenfunctions are used which explains about 96\% variability of the total variation. We apply the same procedure as as done in the case of the previous data set to detect the changes. Table \ref{Table3} shows the outcomes of the test. The functional data representation of the complete data and segment wise mean functions  are  shown in Figure \ref{fig:global} which reflects the prominent changes around the mentioned  period of year. From the analysis of the second the data set we clearly observe that the global temperature is changing (more specifically increasing) significantly over the period of time. 
%Further we analyze the same kind of temperature data, that is monthly temperature data for norther hemisphere and southern hemisphere to see temperature of which hemisphere is effecting the global temperature most. It is seen in Table ??? that the temperature of the northern hemisphere has increased over the said period of time more prominently than the other part, from which one may draw the conclusion that the effect of global warming is more prominent in northern hemisphere than southern hemisphere. 

\section{Discussions and conclusions }\label{sec6}
In this paper we have proposed a new test for testing the existence of a change point in a given sequence of independent functional data. It is shown that  the null distribution of proposed test is asymptotically pivotal. We have proven that under the null hypothesis the distribution of the test statistics is a functional  of the sum of squares of Brownian bridges. Moreover, it has been established that under alternative hypothesis of single change point the power of the proposed test goes to unity when sample sizes increases to infinity.  While developing the test statistic we have proposed an alternative  estimator  of the covariance kernel, which is  not only a consistent estimator of the true covariance kernel under the null hypothesis but also it has lesser  bias than the existing usual estimate of covariance kernel under the alternative hypothesis. In fact it is successfully shown that even under the alternative hypothesis, if the data is divided at the true point of change then our estimate has zero asymptotic bias whereas the existing estimate of covariance kernel mostly used in change point literature in functional data has a constant asymptotic bias. Because of the fact that our used estimate of covariance kernel has a smaller bias than the existing one under any circumstances, we are able to show that our test has greater power than the existing one for testing the presence of change point in a given sequence of functional data. The  extensive simulation studies support such a claim also. Specially when the data size is not very big then our method outperforms the existing one with a great margin. 
\par 
We have used our method in two real data to see the performance of our test in practice. One of these data is central England temperature which is also used in \cite{Berkes_2009}, and the other one is the global temperature data. In case of first data, it is seen that our method and the method of \cite{Berkes_2009} both, have pointed four changes in the data sequence. Two time points have exactly matched for two methods, namely 1850 and 1926. For two other change points two methods differ marginally. \cite{Berkes_2009} has detected 1808 as possible change point whereas our method detected 1810 as possible change point. For the other one \cite{Berkes_2009} detected 1993 as a possible change point and our method indicated 1989 as a possible change point. We have plotted the mean function for each of the different segments which clearly shows an upward trend in the mean temperature over the said periods. The mean curves of different time segments are very similar to that of \cite{Berkes_2009} which make sure the little observed difference in change points among two methods in this particular real data are not major. For the second data, which is global monthly temperature data from 1850 to 2012, is analyzed based on our method. It is found that there exists three change points around 1933, 1986 and 1996. The analysis of global temperature in terms of finding change points will help the scientists working on the global temperature. It clearly shows that in last three decades the temperature has increased significantly over the past. 
\par 
To conclude  we evince  that the proposed method has asymptotic null pivotal distribution with greater  power than the existing method for testing the presence of change in a sequence of functional data and hence can be used in practice with more confidence.

\section*{Acknowledgment}
The authors are thankful to British Atmospheric Data Centre and carbon dioxide information analysis center for real data. 
The Daily Central England Temperature data has been taken from \cite{British_temp}
and monthly global average anomaly of temperatures is taken from 
\cite{global_temp}. 
\section{Appendix}\label{sec7}

Proof of the Theorem \ref{thm1}\/: Define $\widehat{\mu}_k(t)=\displaystyle\frac{1}{k}\sum_{i=1}^{k}X_i(t)$  and 
$\widetilde{\mu}_k(t)=\displaystyle\frac{1}{N-k}\sum_{i=k+1}^{N}X_i(t)$  for some $k=[Nu]$ and $k^*=[N\theta]$ to express  the estimated covariance kernel as 
\begin{eqnarray}
\widehat c_u(t,s)=\frac{1}{N}\left[\sum_{i=1}^k\{X_i(t)-\widehat{\mu}_k(t)\}\{X_i(s)-\widehat{\mu}_k(s)\}\right. 
+\left.\sum_{i=k+1}^N\{X_i(t)-\widetilde{\mu}_k(t)\}\{X_i(s)-\widetilde{\mu}_k(s)\}\right] \nonumber
\end{eqnarray}
It immediately  gives 
\begin{eqnarray}
\widehat c_u(t,s)=&&\frac{1}{N}\sum_{i=1}^N
\{X_i(t)-  \widehat{\mu}_N(t)\}\{X_i(s)-  \widehat{\mu}_N(s)\} 
-\frac{k}{N} \{\widehat{\mu}_k(t)-\widehat{\mu}_N(t)\}\{\widehat{\mu}_k(s)-\widehat{\mu}_N(s)\} \nonumber\\
&-&\frac{k}{N} \{\widetilde{\mu}_k(t)-\widehat{\mu}_N(t)\}\{\widetilde{\mu}_k(s)-\widehat{\mu}_N(s)\} \nonumber
\end{eqnarray}

For $k\leq k^*$, note that 
$$\widehat{\mu}_k(t)= \widehat{\overline{Y}}_k(t)+\mu_1(t)\mbox{~~where,~~} \widehat{\overline{Y}}_k(t)= \displaystyle\frac{1}{k}\sum_{i=1}^{k}Y_i(t), $$
$$\widetilde{\mu}_k(t)= \widetilde{\overline{Y}}_k(t)+\mu_2(t)+\left(\frac{k^*-k}{N-k}\right)\Delta(t)\mbox{~~where,~~} \widetilde{\overline{Y}}_k(t)= \displaystyle\frac{1}{k}\sum_{i=k+1}^{N}Y_i(t),  $$
and 
$$\widehat{\mu}_N(t)= \widehat{\overline{Y}}_N(t)+\left(\frac{k^*}{N}\right)\mu_1(t)+\left(\frac{N-k^*}{N}\right)\mu_2(t)$$

Now observe that 
$$\widehat{\mu}_k(t)-\widehat{\mu}_N(t)=\widehat{\overline{Y}}_k(t)-\widehat{\overline{Y}}_N(t)+\left(1-\frac{k^*}{N}\right)\Delta(t)$$
and 
$$\widetilde{\mu}_k(t)-\widehat{\mu}_N(t)=\widetilde{\overline{Y}}_k(t)-\widehat{\overline{Y}}_N(t)-\frac{k(N-k^*)}{(N-k)N}~\Delta(t)$$
to get  the following deductions,
\begin{eqnarray}
\widehat c_u(t,s)=&&\frac{1}{N}\sum_{i=1}^N
\{X_i(t)-  \widehat{\mu}_N(t)\}\{X_i(s)-  \widehat{\mu}_N(s)\} -\Delta(t)\Delta(s)\left(\frac{N-k^*}{N}\right)^2\left(\frac{k}{N-k}\right)\nonumber\\
&-&\frac{k}{N} \{\widehat{\overline{Y}}_k(t)-\widehat{\overline{Y}}_N(t)\}\{\widehat{\overline{Y}}_k(s)-\widehat{\overline{Y}}_N(s)\} -\left(1-\frac{k}{N}\right) \{\widetilde{\overline{Y}}_k(t)-\widehat{\overline{Y}}_N(t)\}\{\widetilde{\overline{Y}}_k(s)-\widehat{\overline{Y}}_N(s)\} \nonumber\\
&-&\frac{k}{N} \left(1-\frac{k^*}{N}\right)\left[\{\widehat{\overline{Y}}_k(t)-\widetilde{\overline{Y}}_k(t)\}\Delta(s)+ \{\widehat{\overline{Y}}_k(s)-\widetilde{\overline{Y}}_k(s)\}\Delta(t)\right] \nonumber
\end{eqnarray}
Again,
\begin{eqnarray}
\widehat c_1(t,s)&=&\frac{1}{N}\sum_{i=1}^N
\{X_i(t)-  \widehat{\mu}_N(t)\}\{X_i(s)-  \widehat{\mu}_N(s)\} \nonumber\\
&=&\frac{1}{N}\sum_{i=1}^N
\{Y_i(t)- \widehat{\overline{Y}}_N(t)\}\{Y_i(s)- \widehat{\overline{Y}}_N(s)\} +\frac{k^*}{N} \left(1-\frac{k^*}{N}\right)\Delta(t)\Delta(s) \nonumber\\
&& + \frac{k^*}{N} \left(1-\frac{k^*}{N}\right)\left[\{\widehat{\overline{Y}}_k(t)-\widetilde{\overline{Y}}_k(t)\}\Delta(s)+ \{\widehat{\overline{Y}}_k(s)-\widetilde{\overline{Y}}_k(s)\}\Delta(t)\right] \nonumber
\end{eqnarray}
gives,
\begin{eqnarray}
\widehat c_u(t,s)&=&\frac{1}{N}\sum_{i=1}^N
\{Y_i(t)- \widehat{\overline{Y}}_N(t)\}\{Y_i(s)- \widehat{\overline{Y}}_N(s)\} +\frac{k^*}{N} \left(1-\frac{k^*}{N}\right)\left[1-\frac{(N-k^*)k}{(N-k)k^*}\right]\Delta(t)\Delta(s) \nonumber\\
&&+\left(1-\frac{k^*}{N}\right)\Delta(s)\left[ \frac{k^*}{N}\{\widehat{\overline{Y}}_{k^*}(t)-\widetilde{\overline{Y}}_{k^*}(t)\}-\frac{k}{N}\{\widehat{\overline{Y}}_k(t)-\widetilde{\overline{Y}}_k(t)\}\right]\nonumber\\
&&+\left(1-\frac{k^*}{N}\right)\Delta(t)\left[ \frac{k^*}{N}\{\widehat{\overline{Y}}_{k^*}(s)-\widetilde{\overline{Y}}_{k^*}(s)\}-\frac{k}{N}\{\widehat{\overline{Y}}_k(s)-\widetilde{\overline{Y}}_k(s)\}\right] \nonumber\\
&&-\ \frac{k}{N}\left(1-\frac{k}{N}\right) \{\widehat{\overline{Y}}_k(t)-\widetilde{\overline{Y}}_k(t)\}\{\widehat{\overline{Y}}_k(s)-\widetilde{\overline{Y}}_k(s)\} \nonumber\\
&\equiv&\frac{1}{N}\sum_{i=1}^N
\{Y_i(t)- \widehat{\overline{Y}}_N(t)\}\{Y_i(s)- \widehat{\overline{Y}}_N(s)\} +\theta(1-\theta)\Delta(t)\Delta(s)~f_{\theta}(u) \nonumber\\
&&+r_1(t,s)+r_2(t,s)+r_3(t,s), \mbox{~~~say} 
\end{eqnarray}

Using the law of large numbers for independent, identically distributed Hilbert-space-valued random
variables (see for example theorem 2.4 of \cite{Bosq_2000}), we obtain 
$$\int_{\tau}\int_{\tau} r_1^2(t,s) dt ds\xrightarrow{P}0 \mbox{~~and ~~}\int_{\tau}\int_{\tau} r_2^2(t,s) dt ds\xrightarrow{P}0 \mbox{~~as~~} N\rightarrow \infty.$$
At the same time using theorem 5.1 of \cite{Inference_book2012} we get 
$$N^2 \int_{\tau}\int_{\tau} r_3^2(t,s) dt ds\xrightarrow{d} \left(\int_{\tau}\Gamma^2(t)dt\right)^2,$$ where $\{\Gamma(t):t\in \tau\}$ is a Gaussian process with $E(\Gamma(t))$ = 0 and $E(\Gamma(t)\Gamma(s))$ = $c(t,s)$, 
which in turn implies that 
$$\int_{\tau}\int_{\tau} r_3^2(t,s) dt ds \xrightarrow{P}0 \mbox{~~as~~} N\rightarrow \infty.$$
These help to conclude that 
$$\int_{\tau}\int_{\tau}[\widehat c_u(t,s)-c_u(t,s)]^2dt ds\xrightarrow{P} 0 \mbox{~~as~~} N\rightarrow \infty.$$
The similar proof holds when $k> k^*$.
It is easy to see that under the  null  hypothesis 
$$\widehat c_u(t,s)\xrightarrow{P} c(t,s) ~~~ \forall u \in(0,1] \mbox{~~as~~} N\rightarrow \infty $$
\hfill $\square$
%----------------------------------------------%
%\begin{theorem}
%	 Under the null hypothesis, for all $u\in (0,1]$
%\begin{equation}
%	R_N(u)=\displaystyle \frac{1}{N}\sum_{l=1}^d\frac{1}{\widehat{\lambda}^{l}_u}\left(\sum_{i=1}^{[Nu]}\widehat\eta_{i,l}(u)-u\sum_{i=1}^{N}\widehat\eta_{i,l}(u)\right)^2\xrightarrow{d}\displaystyle \sum_{l=1}^{d}B_l^2(u)
%\end{equation}
%where $ B_1( . ), B_2( . ), ....., B_d( . )$ denote independent standard Brownian bridges over $[0,1]$
%\end{theorem}
\\
\noindent
Proof of Theorem \ref{thm2}\/:  

The proof follows from the Theorem \ref{thm1}, Corollary \ref{cor1} and the proof of Theorem 6.1 of \cite{Inference_book2012}. 

%\newpage
\begin{scriptsize}
 \bibliographystyle{abbrvnat}
\bibliography{buddha_bib.bib}
\end{scriptsize}
\newpage
\begin{scriptsize}
	
\begin{table}[h]
	% \begin{threeparttable}
	
	\centering
	\caption{\scriptsize Power comparison of two tests with test statistics $S_{N,d}$ and $H_{N,d}$ for different $k^*$}
	%\footnote{$^*$ The values are reported from the tables provided by  Berkes et.\ al (2009).}
	\begin{tabular}{cccccccccc}\hline
		\multicolumn{1}{c}{$N=100, d=3$} & & \multicolumn{2}{c}{BM,BM+t}& & \multicolumn{2}{c}{BM,BM+$sin(t)$}& & \multicolumn{2}{c}{BB,BB+$0.8(1-t)t$}\\ 
		\cline{1-1} \cline{3-4} \cline{6-7} \cline{9-10}  
		$k^*$ & & $S_{N,d}$ &$H_{N,d}$& & $S_{N,d}$ &$H_{N,d}$&& $S_{N,d}$ &$H_{N,d}$\\
		\hline 
		0	&	&	4.6$^*$	&	5.5	&	&	4.6$^*$	&	5.5	&	&	4.6$^*$	&	5.0	\\
		15	&	&	36.3	&	39.9	&	&	30.9	&	35.4	&	&	11.2	&	12.9	\\
		20	&	&	57.3	&	62.0	&	&	44.6	&	49.5	&	&	15.3	&	16.5	\\
		25	&	&	72.0	&	75.6	&	&	61.2	&	64.7	&	&	19.5	&	21.3	\\
		35	&	&	92.9	&	94.2	&	&	80.1	&	83.4	&	&	28.0	&	31.8	\\
		50	&	&	94.9$^*$	&	95.8	&	&	88.0$^*$	&	90.1	&	&	34.7	&	37.4	\\
		65	&	&	91.0	&	92.9	&	&	81.5	&	83.7	&	&	31.1	&	33.9	\\
		75	&	&	74.3	&	78.1	&	&	59.0	&	64.4	&	&	21.9	&	23.8	\\
		80	&	&	58.8	&	64.3	&	&	46.1	&	50.2	&	&	13.7	&	16.1	\\
		85	&	&	36.4	&	40.1	&	&	27.8	&	33.0	&	&	12.9	&	14.1	\\
\hline 
\hline 
	\end{tabular} 
	
	% \end{threeparttable}
	\label{Table1}
\end{table}
$^*$ The values are reported from the tables provided by  Berkes et.\ al (2009,  Table 3 ).
\vspace{1 cm}
\begin{table}[h]
	% \begin{threeparttable}
	
	\centering
	\caption{\scriptsize  Comparisons of the performance of  $S_{N,d}$ and $H_{N,d}$  for UK temperature data}
	%\footnote{$^*$ The values are reported from the tables provided by  Berkes et.\ al (2009).}
	\begin{tabular}{ccccccccc}\hline
		\multicolumn{4}{c}{Performance of  $S_{N,d}^*$ } & & \multicolumn{4}{c}{ Performance of  $H_{N,d}$}\\
		\cline{1-4} \cline{6-9} 
		Year & Observed &  Obtained  & Estimated & & Year & Observed &  Obtained  & Estimated\\
		Segment& $S_{N,d}$& P-value & Change point & & Segment& $H_{N,d}$& P-value &  Change point\\
		\cline{1-4} \cline{6-9} 
		1780-2007& 8.020593& 0.00000 & 1926& & 1780-2007& 9.820036 & 0.00000& 1926\\
		1780-1925& 3.252796 & 0.00088& 1808& & 1780-1926& 3.764348& 0.00011& 1850\\
		1808-1925& 2.351132 & 0.02322& 1850 & & 1780-1850 & 2.403308&  0.01900& 1810 \\
		1926-2007&  2.311151& 0.02643 & 1993 && 1927-2007 & 2.649414 & 0.00797 & 1989\\
	\hline 
		\hline 
	\end{tabular} 
	\label{Table2}
	% \end{threeparttable}
\end{table}

$^*$ The values are reported from the tables provided by  Berkes et.\ al (2009 , Table 4).
\vspace{1 cm}
\begin{table}[h]
	% \begin{threeparttable}
	
	\centering
	\caption{\scriptsize  Change points for average anomaly global temperature data}
	%\footnote{$^*$ The values are reported from the tables provided by  Berkes et.\ al (2009).}
	\begin{tabular}{cccc}\hline
		\multicolumn{4}{c}{Performance of  $H_{N,d}$ } \\
		\cline{1-4}
		Year  Segment& Observed $H_{N,d}$ &  Obtained   P-value& Estimated  Change point  \\
	%	Segment& $H_{N,d}$& P-value & Change point \\
		
		\cline{1-4}
		1850-2012& 23.63304& 0.00000 & 1933\\
		1934-2012& 13.46585 & 0.00000& 1986\\
		1987-2012&  4.34103& 0.00000& 1996\\
		\hline 

	\end{tabular} 
	\label{Table3}
	% \end{threeparttable}
\end{table}
\end{scriptsize}
\newpage

\newpage

\end{document}